\documentclass[lettersize,journal]{IEEEtran}

\usepackage{cite}
\usepackage{amsmath,amsfonts}
\usepackage{algorithmic}
\usepackage{algorithm}
\usepackage{array}
\usepackage[caption=false,font=normalsize,labelfont=sf,textfont=sf]{subfig}
\usepackage{textcomp}
\usepackage{stfloats}
\usepackage{url}
\usepackage{verbatim}
\usepackage{graphicx}
\usepackage{adjustbox}

\usepackage{cite}
\usepackage{amsmath,amssymb,amsfonts}
\usepackage{algorithmic}
\usepackage{graphicx}
\usepackage{textcomp}
\usepackage{xcolor}
\usepackage{graphicx}
\usepackage{mathtools}
\usepackage{caption}
\usepackage[utf8]{inputenc}
\DeclareSymbolFont{matha}{OML}{txmi}{m}{it}
\DeclareMathSymbol{\varv}{\mathord}{matha}{118}
\usepackage{stackengine}

\usepackage[english]{babel}
\usepackage{amsthm}
\usepackage{soul,color}
\usepackage{algorithm}
\usepackage{setspace}
\newtheorem{remark}{Remark}
\usepackage{lipsum}
\usepackage[figurename=Fig.]{caption}

\usepackage{hyperref}
\begin{document}

\title{Parallel Coding for Orthogonal Delay-Doppler Division Multiplexing}

\author{Qi Li,~\IEEEmembership{Student Member,~IEEE,}
        Jinhong Yuan,~\IEEEmembership{Fellow,~IEEE,}
        Min Qiu,~\IEEEmembership{Member,~IEEE,}
\thanks{
The work was supported in part by the Australian Research Council (ARC) Discovery Project under Grant DP220103596, and in part by the ARC Linkage Project under Grant LP200301482. This work has been presented in part at the 2023 IEEE International Conference on Communications Workshops, Rome, Italy \cite{Li2023ParallelFeedback}.

Qi Li, Jinhong Yuan and Min Qiu are with the School of Electrical Engineering and Telecommunications, The University of New South Wales, Sydney, Australia (e-mail: oliver.li1@unsw.edu.au, j.yuan@unsw.edu.au, min.qiu@unsw.edu.au, yixuan.xie@unsw.edu.au).\par

}%

}



\maketitle

\begin{abstract}
This paper proposes a novel parallel coding transmission strategy and an iterative detection and decoding receiver signal processing technique for orthogonal delay-Doppler division multiplexing (ODDM) modulation. Specifically, the proposed approach employs a parallel channel encoding (PCE) scheme that consists of multiple short-length codewords for each delay-Doppler multicarrier (DDMC) symbol. Building upon such a PCE transmission framework, we then introduce an iterative detection and decoding algorithm incorporating a successive decoding feedback (SDF) technique, which enables instant information exchange between the detector and decoder for each DDMC symbol. To characterize the error performance of the proposed scheme, we perform density evolution analysis considering the finite blocklength effects. Our analysis results, coupled with extensive simulations, demonstrate that the proposed PCE scheme with the SDF algorithm not only showcases a better overall performance but also requires much less decoding complexity to implement, compared to the conventional benchmark scheme that relies on a single long channel code for coding the entire ODDM frame.
\end{abstract}


\section{Introduction}
\IEEEPARstart In the upcoming era of wireless communications, there is an anticipation of accommodating diverse high-mobility communication scenarios characterized by a high Doppler frequency shift. This shift poses a new challenge to existing modulation schemes, such as the orthogonal frequency division multiplexing (OFDM) modulation. To mitigate the adverse effects of Doppler and enhance overall communication system performance, a novel modulation scheme called orthogonal time frequency space (OTFS), proposed by \cite{Hadani2017OrthogonalModulation}, has demonstrated effectiveness at communicating in doubly-selective wireless channels. OTFS is a novel two-dimensional (2D) modulation scheme in which symbols are modulated on the delay-Doppler (DD) domain that can better couple with the quasi time-invariant DD channel. Despite such coupling, the bi-orthogonality robust pulse assumed in OTFS cannot be realized in practice \cite{Raviteja2019PracticalOTFS}. As a workaround solution, the TF domain rectangular pulse is used to form a signal structure based on OFDM, which renders OTFS compatible with the existing modulation system \cite{Raviteja2019PracticalOTFS}. {Regarding} signal recovery at the receiver, this interim pulse solution presents challenges in accurately recovering the transmitted signals. {The complicated interference pattern} is due to the imperfect coupling of the channel with the non-ideal pulses, resulting in significant interference. Conversely, while most research on OTFS presupposes the use of rectangular pulse shaping, it encounters issues such as high out-of-band emission (OOBE) and significant intersymbol interference (ISI) when a necessary and practical band-pass filter is applied at the receiver \cite{Shen2022ErrorReceivers}. To address this fundamental {pulse-shaping} issue, the authors in \cite{Lin2022OrthogonalModulation} proposed an orthogonal delay-Doppler division multiplexing (ODDM) modulation. {Unlike} OTFS modulation, ODDM uses a square-root Nyquist pulse train to achieve sufficient orthogonality on the DD plane with fine resolutions \cite{Lin2022OrthogonalModulation}, \cite{Lin2023Multi-CarrierDomain}. \par

Among the variant OTFS systems, the zero-padded OTFS proposed in \cite{Thaj2020LowSystems} allows extra detection freedom on the receiver side. The zeros (null symbols) added in the DD domain divide the whole signal frame into independent blocks in parallel. Such property leads to the development of some low complexity linear detection methods such as maximal ratio combining (MRC) in \cite{Thaj2020LowSystems} and successive interference cancellation based on minimum mean squared error (SIC-MMSE) detection in \cite{Li2022IterativeModulation}. In \cite{Li2022IterativeModulation}, the proposed detection recovers the signals using MMSE filtering in the time domain with hard decision; however, the hard decision has limited the detection performance. On the other hand, non-linear detectors such as maximum \emph{a posteriori} (MAP) can achieve optimal performance, but this is impractical due to their exceedingly high complexity \cite{LR1,Murali2018OnChannels}. To reduce the complexity imposed by MAP, several message passing (MP) based detection variants were proposed. In \cite{LR3}, the authors proposed a message passing algorithm (MPA) for joint interference cancellation and signal detection with reduced computational complexity. However, it was then pointed out in \cite{LR4} that MPA is very likely to converge to a locally optimal point in the loopy factor graph. A variational Bayes (VB) approach was then proposed by authors in \cite{LR4}, which can lead to a guaranteed convergence with significant performance gains. Comparisons on the performance between different detectors for ODDM were conducted in \cite{Huang2023PerformanceEstimation}. \par
Despite the progress of the above research, the design and analysis of channel coding schemes for OTFS or ODDM have received less attention. Most works adopt the traditional single long channel codes such as low-density parity-check (LDPC) codes \cite{Das2020PerformanceOTFS}, \cite{Long2019LowOTFS} or convolutional codes \cite{Li2021PerformanceChannels} for coding the entire OTFS frame, which typically has a separate detection and decoding module. For coded ODDM, the work in \cite{Lin2022OrthogonalModulation}, however, still needs to emphasize more the coding aspect of the system and uses the traditional long channel code for the entire ODDM frame. The use of long codewords undeniably introduces additional computational complexity stemming from the complicated decoding process and lack of seamless integration between detection and decoding, thus adversely affecting performance. However, some work proposed utilizing multiple codewords for coding over the entire OTFS frame \cite{Thaj2020LowSystems}. Despite an increase in the number of codewords and a decrease in codeword length, the detection and decoding processes remain segregated. Recently, the authors in \cite{Wang2024ExploitingApproach} integrated the detector and decoder for LDPC-enhanced Delay-Doppler multicarrier systems into a unified framework using a factor graph to exploit the joint sparsity of the delay-Doppler channel and the LDPC parity-check matrix. In contrast, this article proposes a parallel-encoded system that focuses not only on reducing complexity through channel sparsity but also on enhancing interference cancellation to improve overall error performance, offering a distinct approach to addressing the challenges in DDMC systems. In addition, the impacts of channel coding on the performance of OTFS or ODDM have not been fully understood. Although \cite{Das2020PerformanceOTFS} investigates the performance of coded OTFS under maximum-likelihood decoding, the characterization on the performance of modern channel codes, e.g., LDPC codes, under iterative decoding over OTFS or ODDM, remains lacking. 

Motivated by the complexity challenges and the absence of integration between detector and decoder in 2D modulation systems such as ODDM, we propose a parallel encoded system that integrates the detection and decoding processes for ODDM systems and introduces an iterative joint detection and decoding receiver for the proposed scheme \cite{Li2023ParallelFeedback}, showcasing its superior performance and reduced complexity. This paper is driven to explore parallel coding for ODDM with successive decoding feedback. As an extension of our previous work \cite{Li2023ParallelFeedback}, this article presents a comprehensive error performance and complexity analysis, along with additional simulation results to further validate our findings. The detailed contributions of this paper are as follows:
\begin{itemize}
\item We introduce the structure of the parallel channel encoded (PCE) ODDM system and the joint detection and decoding algorithm. Specifically, the PCE scheme encodes each delay-Doppler multicarrier symbol with independent codewords. This encoding design enables a simplified yet robust joint detection and decoding algorithm at the receiver, which allows instant information exchange between the detector and decoder for each row signal. This is in contrast to the traditional single codeword encoded (SCE) system in which the decoding is performed only after the detection of the whole frame. Based on the PCE, we then derive this soft interference cancellation based SIC-MMSE detection and propose an iterative joint detection and decoding algorithm with successive decoding feedback (SDF).
\item We conduct a system performance analysis using density evolution and finite blocklength scaling law tailored to specific code designs. We first characterize the detector output as Gaussian or symmetric Gaussian mixtures. Second, we utilize the density evolution (DE) to track the information exchange between bit nodes and check nodes in the LDPC decoding process. Building upon the decoding threshold analysis, we further analyze the error probability of LDPC coded ODDM by accounting for the finite blocklength effects. This allows a fair comparison of the theoretical performance between the proposed PCE scheme and the benchmark SCE ODDM scheme.
\item We analyze the detection and decoding complexity for PCE and benchmark SCE scheme. We also provide the simulation results employing off-the-shelf LDPC codes with various code rates and modulation orders. Our analysis and simulation results show that our proposed PCE design with short codes {improves} the error rate and reduces the complexity needed to achieve a comparable error rate compared to the benchmark scheme SCE with a single long code.
\end{itemize}

The rest of the paper is organized as follows: Sec. \ref{sec:model} introduces the modulation and demodulation process for ZP-ODDM, and discusses the input-output relation and effective channel matrix in the time domain. In Sec. \ref{sec:prop}, we then introduce the iterative cross-domain SIC-MMSE signal detection ZP-ODDM. In the same section, we also introduce our proposed PCE scheme and the joint detection and decoding algorithm with SDF for the proposed PCE ODDM. In Sec. \ref{sec:analysis}, we analyze the error performance of our proposed PCE and traditional SCE scheme by utilizing density evolution analysis and finite blocklength scaling law, and its complexity. Our error analysis can closely match the simulation results, which will be shown in the next section. Finally, Sec. \ref{sec:sim} provides extensive simulation results data for different modulation orders and code rates, and a clear performance superiority of our proposed PCE ODDM scheme can be observed. Last but not least, Sec. \ref{sec:con} concludes our findings.

\subsection*{Notations:}
We will use the following notations throughout this paper: $x$, $\bf x$ and $\bf X$ represent scalar, vector and matrix. ${\bf X}^{T}$ and ${\bf X}^{\dag}$ represent transpose and conjugate transpose. $\bf 0$, ${\bf I}_{N}$, ${\bf F}_{N}$ and ${\bf F}_{N}^{\dag}$ are zero matrix, identity matrix with order $N$, $N$-point normalized discrete Fourier transform (DFT) matrix and $N$-point normalized inverse discrete Fourier transform (IDFT) matrix, respectively. Let $\circledast$, $\otimes$ and $\circ$ denote circular convolution, Kronecker product and Hadamard product. $\text{vec}(\bf X)$ and $\text{vec}_{N,M}^{-1}(\bf x)$ represent column-wise vectorization of matrix $\bf X$ and matrix formed by folding a vector $\bf x$ into a $N\times M$ matrix, respectively. The set of $N\times M$ dimensional matrices with complex entries are denoted by $\mathbb {C}^{N \times M}$. ${\bf H}_{n,m}$ denotes the $m$-th sub-channel matrix at the $n$-th signal block and ${\bf H}_{n,m}[:,l]$ denotes the $l$-th column vector in matrix ${\bf H}_{n,m}$.

\section{ODDM System Model}\label{sec:model}
We consider a zero-padded (ZP) ODDM system spanning over the time interval $(-Q\frac{T}{M}, NT+(Q-1)\frac{T}{M})$, where $Q$ is an integer
and $2Q \ll M$, $M$ is the number of delay-Doppler domain staggered multicarrier (DDMC) symbols, $N$ is the number of subcarriers; and a frequency spectrum limited to $(-\frac{M}{2T},\frac{M}{2T})$ approximately. The ODDM has a delay resolution of $\frac{T}{M}$, and a Doppler resolution of $\frac{1}{NT}$ equivalent to the DDMC subcarrier spacing. \par
Let $x[m,n]$, where $m=\{0,\ldots,M-L-1\}$ and $n=\{0,\ldots,N-1\}$, denote the $(m,n)$-th modulated $\mathcal{Q}$-QAM data symbol that is arranged on the two-dimensional DD domain grid ${\bf X}_{\rm DD} \in \mathbb {C}^{M \times N}$, where $N$ and $M$ are also the number of Doppler bins and delay bins per ODDM frame, respectively. In ZP-ODDM, the zero paddings or null symbols are added to the last $L$ rows of ${\bf X}_{\rm DD}$, where $L$ corresponds to the largest delay index of the channel response. There are quite a few advantages of using zero paddings; for instance, it allows parallel processing for each column symbol vector either in the DD domain or time domain. Besides, it can also facilitate the embedded channel estimation approach such as the pilot-based scheme in \cite{Huang2023PerformanceEstimation}, \cite{Raviteja2019EmbeddedChannels}. Nevertheless, the ultimate ODDM waveform can be regarded as staggering $M$ ODDM symbols in the time domain. The $m$-th ODDM symbol ($m$-th row in DD domain) waveform $s_{m}(t)$ is sample-wise generated using a truncated square-root Nyquist pulse $a(t)$ by \cite{Lin2022OrthogonalModulation}

\begin{align}
{s_{m}(t)}=\sum_{\dot{n}=0}^{N-1}\sum_{n=0}^{N-1}x[m,n]e^{j2\pi\frac{\dot{n}n}{N}}a(t-\dot{n}T), \quad \label{eq200}
\end{align}
where $\dot{n}$ and $n$ are the indices of time and frequency (Doppler), respectively, $T$ is the adjacent Nyquist pulse interval in the $m$-th ODDM symbol before upsampling, and $a(t)$ is the square-root Nyquist pulse. This sample-wise pulse shaped waveform can be equivalently implemented by adding up $a(t)$ and form $u(t) = \sum\limits_{\dot{n}=0}^{N-1}a(t-\dot{n}T)$. With $u(t)$, the $m$-th time domain ODDM symbol can also be represented as \cite{Lin2022OrthogonalModulation}
\begin{align}
{\tilde{s}_{m}(t)} = \sum_{n=0}^{N-1}x[m,n]e^{j2\pi\frac{nt}{NT}}u(t),
\end{align}
After generating the continuous waveform for each $m$-th ODDM symbol, the staggering of the $M$ multi-carrier symbols is then performed \cite{Lin2022OrthogonalModulation}
\begin{align}
{{s}(t)} = \sum_{m=0}^{M-1}\sum_{n=0}^{N-1}x[m,n]e^{j2\pi\frac{n}{NT}(t-m\frac{T}{M})}u\Bigl(t-m\frac{T}{M}\Bigl),
\end{align}
where ${{s}(t)}$ is the ultimate ODDM time domain waveform, and $u(t) = \sum\limits_{\dot{n}=0}^{N-1}a(t-\dot{n}T)$. Then, the CP prepended waveform ${{s_{cp}}(t)}$ will be sent into the doubly-selective wireless channel.\par
Equivalently, the aforementioned modulation process can be implemented digitally first and then pulse shaped. Specifically, the $m$-th delay/row DD domain symbol vector first goes through an $N$ point IFFT and forms the discrete sequence of the $m$-th ODDM symbol in the delay time domain. Note that the zero paddings added at the last $L$ rows after IFFT are still zeros. Then, each $m$-th delay time domain ODDM discrete symbol is upsampled by a factor of $M$ and then staggered by an interval $\frac{T}{M}$. Finally, the staggered discrete symbols are sample-wise pulse shaped by $a(t)$ and eventually arrive at ${{s}(t)}$ {equivalently}.\par
A high-mobility wireless channel can be represented by a linear time-variant (LTV) system, a.k.a. the doubly-selective channel \cite{Bello1963CharacterizationChannels}. A physically meaningful representation of the LTV wireless channel is based on the time delays and Doppler frequency shifts \cite{Bello1963CharacterizationChannels}. Let $P$ denote the number of resolvable paths in the channel, and the DD domain channel can be represented as \cite{Bello1963CharacterizationChannels} \begin{align} h(\tau, \nu) = \sum _{p=1}^{P} h_p \delta (\tau -\tau _p) \delta (\nu -\nu _p), \label{eq26}\end{align}
where $\tau _p$, $\nu _p$ and $h_p$ are the delay, Doppler shift, and attenuation factor for the $p$-th path, respectively \cite{Bello1963CharacterizationChannels}.

\begin{figure}[!t]
\includegraphics[width=\linewidth]{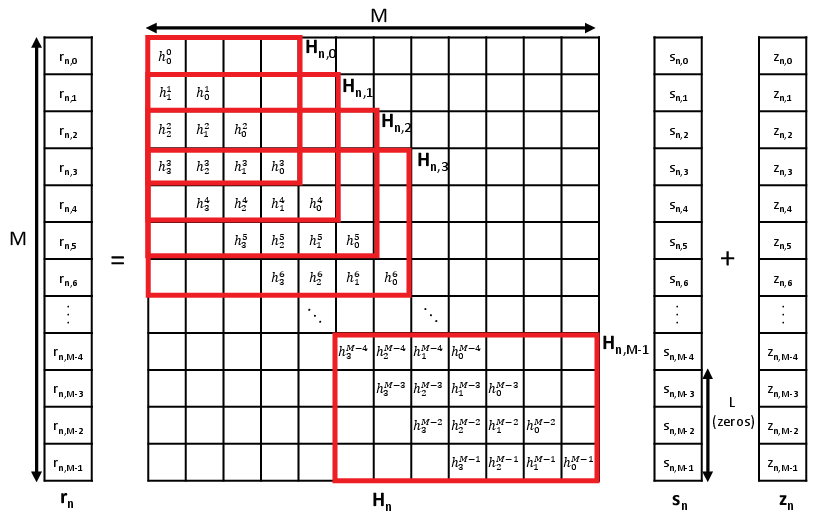}
\caption{Time domain input-output relation for the $n$-th block with $L$ = 3}
\label{f:fig2}
\end{figure}
On the receiver side, the received ODDM signal distorted by the doubly-selective channel can be expressed as
\begin{align}
{{r}(t)} = \sum_{p=1}^{P}h_{p}s(t-\tau_{p})e^{j2\pi v_{p}(t-\tau_{p})}+z(t),\label{eq202}
\end{align}
where $P$ is the number of paths, $h_{p}$, $\tau_{p}$ and $v_{p}$ are the channel coefficient, delay, and Doppler for the $p$-th path, respectively. The received signal $r(t)$ is then sent into the same matched filtering $u(t)$ and the $m$-th ODDM symbol at the $n$-th subcarrier can then be expressed as \cite{Lin2022OrthogonalModulation}
\begin{align}
y[m,n] = \int r(t)u\Bigl(t-m\frac{T}{M}\Bigl)e^{-j2\pi\frac{n}{NT}(t-m\frac{T}{M})}dt.\label{eq203}
\end{align}
Here, we have briefly introduced the zero-padded ODDM modulation. Readers are encouraged to find more details on ODDM modulation in \cite{Lin2022OrthogonalModulation}.\par

\begin{figure*}[!t]
\centering
\includegraphics[width=\linewidth]{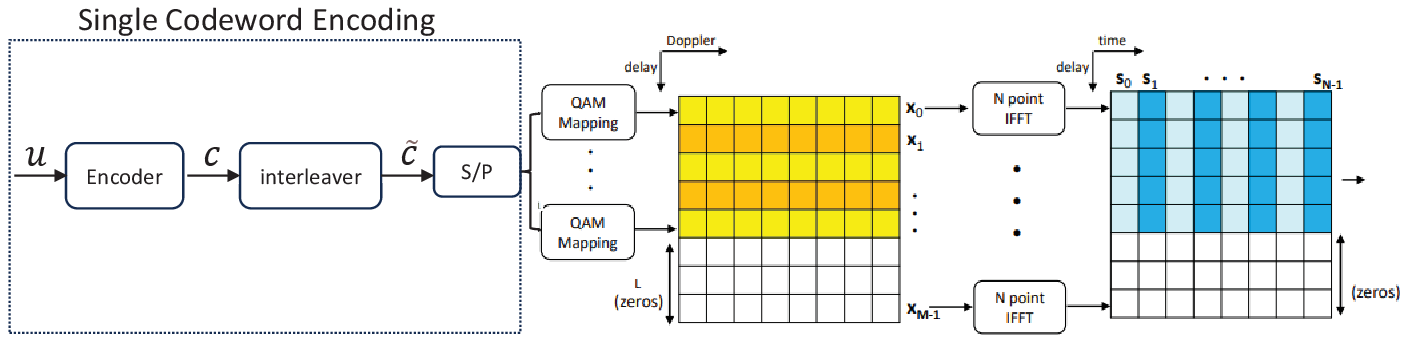}
\captionsetup{justification=centering}
\caption{Single Codeword Encoding (SCE) for ZP-ODDM}
\label{f:fig3}
\end{figure*}

To lay the groundwork for the proposed signal processing algorithm, we first provide the input-output relationship for the staggered length-M symbol vector in the time domain. Processing signals in the time domain is advantageous because Doppler shifts often occur off-grid (fractional Doppler), which leads to a more intricate interference pattern in the DD domain. In contrast, while the time-domain channel cannot fully resolve Doppler diversity, it offers a simpler interference structure for processing.  Furthermore, the insertion of ZP ensures that each length-$M$ symbol vector from a given column in the time domain transmitted signal does not interfere with the $N-1$ symbol vectors from other columns. Denote the $n$-th transmitted staggered ODDM symbol vector in the time domain by ${\bf s}_{n}$ (also shown at the start of each column in Fig. \ref{f:fig3}) and the corresponding $n$-th received symbol vector at the $n$-th time slot by ${\bf r}_{n}$, i.e. ${\bf s}_{n} = [s_{n,0},s_{n,1},\ldots,s_{n,M-1}]^
{T}\in \mathbb {C}^{M\times 1}$ and ${\bf r}_{n} = [r_{n,0},r_{n,1},\ldots,r_{n,M-1}]^{T}\in \mathbb {C}^{M\times 1}$. The relation between ${\bf s}_{n}$ and ${\bf r}_{n}$ can be expressed as
\begin{align}
{{\bf r}_{n}}={{\bf H}_{n}}\cdot {{\bf s}_{n}}+{{\bf z}_{n}}, \label{eq234}
\end{align}
where ${\bf H}_{n}$ is the time domain channel for the $n$-th time domain staggered transmit symbol vector ${\bf s}_{n}$, and ${{\bf z}_{n}}$ is the time domain noise vector. The structure of ${\bf H}_{n}$ is shown in Fig. \ref{f:fig2}. Again, due to the insertion of zero paddings in the DD domain grid, each time domain signal block ${\bf s}_{n}$ does not interfere with any other blocks ${\bf s}_{n’}, \forall n’\neq n$, enabling parallel signal processing for all $N$ blocks simultaneously. For each ${{\bf H}_{n}}$, the entire matrix ${{\bf H}_{n}}$ is partitioned into sub-channel matrices, where the sub-channel ${{\bf H}_{n,m}}$ for $s_{n,m}$ has a maximum size of $L+1$ by $2L+1$. This sub-channel relation can be utilized to simplify the detection process using a SIC-MMSE filter. In the following, we first derive the soft interference cancellation-based MMSE detection, followed by the introduction of the turbo iterative detection and decoding with successive decoding feedback (SDF) for the proposed ODDM frame structure.

\section{Proposed Coding Scheme for ODDM}\label{sec:prop}

\begin{figure*}[tbp]
\centering
\includegraphics[width=\linewidth]{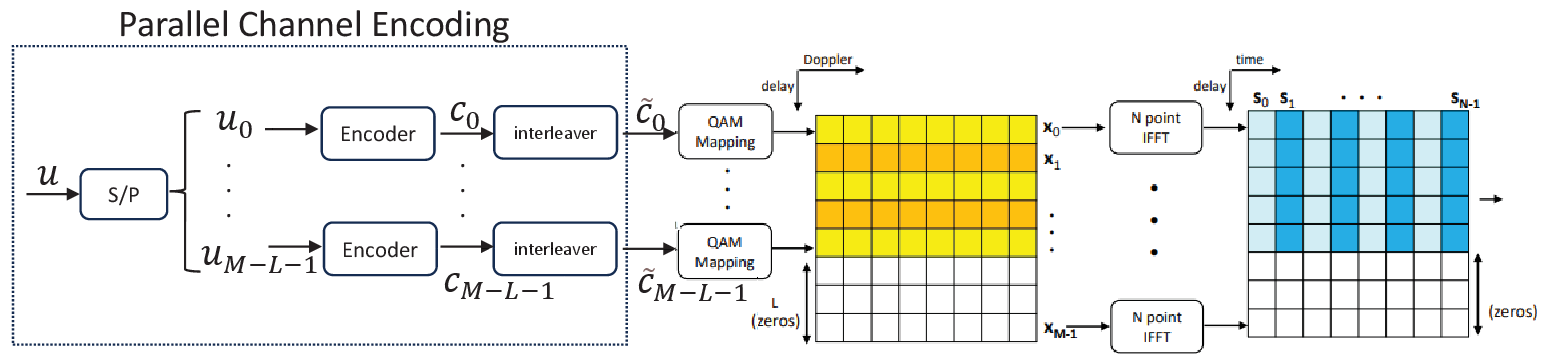}
\captionsetup{justification=centering}
\caption{Parrallel Channel Encoding (PCE) for ZP-ODDM}
\label{f:fig6}
\end{figure*}

\subsection{SCE Scheme for ODDM}\label{sec:sce}
Before we proceed, we first present a baseline scheme that is commonly used in the literature of OTFS and ODDM \cite{Lin2022OrthogonalModulation}, \cite{Das2020PerformanceOTFS}, \cite{Long2019LowOTFS}. As shown in Fig. \ref{f:fig3}, the source information bits ${\bf u}$ is directly encoded and becomes a codeword ${\bf c}$. The codeword length in this case is $n = (M-L)\times N\times \log_{2}|\mathcal{Q}|$. Then, the codeword ${\bf c}$ is interleaved becoming ${\bf \tilde{c}}_{m}$ and converted into parallel sub-codewords of ${\bf \tilde{c}}_{0},{\bf \tilde{c}}_{1},\ldots,{\bf \tilde{c}}_{M-L-1}$ and the $m$-th subcodeword is mapped to the $m$-th row of 2D delay-Doppler domain data symbol. Once all the symbols are formed and arranged on the two-dimensional DD domain grid ${\bf X}_{\rm DD} \in \mathbb {C}^{M \times N}$, the standard ODDM modulation is then performed following Sec. \ref{sec:model}. It is evident that all the symbols in each ODDM frame originate from the bits in a single codeword. Therefore, the detector and decoder remain segregated and do not have a close interaction. In other words, the decoding can only commence after the detection of the whole ODDM frame is finished. However, by designing the codeword uniquely to match the 2D structure of the data symbol, the detector and decoder can interact more frequently, and this is the transmission strategy we will be introducing in Sec. \ref{subsec:PEC}. Before introducing that, we first review an iterative detection signal processing strategy for ZP-ODDM.
\subsection{Iterative SIC-MMSE Detection for ZP-ODDM}\label{subsec:MMSE}
In this section, we briefly present the DD and time cross-domain SIC-MMSE detection {that} will be used for detecting the benchmark SCE scheme in Sec. \ref{sec:sce} based on the work in \cite{Li2024LowModulation}. In brief, the received signals undergo processing one by one through soft interference cancellation and MMSE filtering in the time domain, followed by decision-making in the DD domain. {The updated soft information in the DD domain continually enhances the estimated symbol's statistics through information exchange with the decoder, leading to reduced error rates.} \par
Let $s_{n,m}$ denote the $m$-th transmit symbol at the $n$-th block where $n \in \{0,\ldots,N-1\}$, $m \in \{0,\ldots,M-L-1\}$, and $\bar{{\bf r}}_{n,m}$ denote the receive signal vector that contains the information of $s_{n,m}$ caused by the multi-path channel diversity. To illustrate this, we provide an example based on Fig. \ref{f:fig2}. The received signal vector that corresponds to the transmitted signal layer $s_{n,0}$ is ${\bar{{\bf r}}_{n,0}} = [{r_{n,0},r_{n,1}, r_{n,2}, r_{n,3}}]^T$, and the sub-channel for the signal layer $s_{n,0}$ is ${\bf H}_{n,0}$, where $L=3$ in this example. In general, consider that we want to detect symbol $s_{n,m}$; first of all, the received symbol vector $\bar{{\bf r}}_{n,m} = [r_{n,m},r_{n,m+1},\ldots,r_{n,m+L}]^T\in \mathbb {C}^{(L+1)\times 1}$ associated with $s_{n,m}$ can be expressed as
\begin{align}
{\bar{{\bf r}}_{n,m}} = \sum_{j=0}^{l'+L}{\bf H}_{n,m}[:,j]{s}_{n,(m'+j)}+{\bf z}_{n,m}, \label{eq30}
\end{align}
where we define $m'\triangleq\max\{m-L,0\}$, $l'\triangleq\min\{m,L\}$, and ${\bf H}_{n,m}[:,j]$ is the $j$-th column in ${\bf H}_{n,m}$. The time domain soft interference cancellation in the $i_{d}$-th iteration can then be written as
\begin{align}
{\hat{\bf r}^{i_{d}}_{n,m}} =& \left(\bar{{\bf r}}_{n,m}-\sum_{j=0}^{l'-1}{\bf H}_{n,m}[:,j]\bar{s}_{n,(m'+j)}^{i_{d}}\right.\nonumber\\&-\left.\sum_{k=l'+1}^{l'+L}{\bf H}_{n,m}[:,k]\bar{s}_{n,(m'+k)}^{i_{d}-1}\right)\nonumber\\=&{\bf H}_{n,m}[:,l']s_{n,m}+{\hat {\bf z}^{i_{d}}_{n,m}}, \label{eq31}
\end{align}
where ${\bar s}^{i_{d}}$ and ${\bar s}^{i_{d}-1}$ are the soft estimates of time domain interference variables detected in the $i_{d}$-th iteration and $(i_{d}-1)$-th iteration, respectively, and ${\hat {\bf z}^{i_{d}}_{n,m}}$ is the noise plus residual interference term given by
\begin{align}
{\hat {\bf z}^{i_{d}}_{n,m}} =& \sum\limits_{j=0}^{l'-1}{\bf H}_{n,m}[:,j]e_{n,(m'+j)}^{i_{d}}+\sum\limits_{k=l'+1}^{l'+L}{\bf H}_{n,m}[:,k]e_{n,(m'+k)}^{i_{d}-1}\nonumber\\&+{\bf z}_{n,m}, \label{eq362}
\end{align} where $e^{i_{d}}_{n,(m'+j)}$ and $e^{i_{d}-1}_{n,(m'+k)}$ are the residual interference left in the current and previous iteration, respectively, and ${\bf z}_{n,m}$ is the AWGN. After soft interference cancellation is performed, the MMSE filter is applied to the interference canceled signal vector ${\hat{\bf r}^{i_{d}}_{n,m}}$. The filter coefficients are calculated by
\begin{align}
{\bf w}^{i_{d}}_{n,m} = {\bf H}^{H}_{n,m}[:,l']\left({\bf H}_{n,m}{\bf V}_{n,m}{\bf H}_{n,m}^{H}+\sigma_{n}^2{\bf I}\right)^{-1}, \label{eq37}
\end{align}
where ${\bf H}_{n,m}[:,l']$ is the $l'$-th column of ${\bf H}_{n,m}$ and $l'$ is defined after (\ref{eq30}). The covariance matrix ${\bf V}_{n,m} = $ {diag}$\{{\mathbb V}\{e^{i_{d}}_{n,(m'+1)}\},\ldots,{\mathbb V}\{e^{i_{d}}_{n,(m'+1'-1)}\}, {E_s}, {\mathbb V}\{e^{i_{d}-1}_{n,(m'+1'+1)}\}\\,\ldots,{\mathbb V}\{e^{i_{d}-1}_{n,(m'+1'+L)}\}\}$, in which ${\mathbb V}\{{\cdot}\}$ denotes the variance and ${\mathbb V}\{e^{i_{d}-1}_{n,(m'+k)}\}=E_s$ for $i_{d}=1$ and $k \in \mathbb \{l+1,\ldots,l'+L\}$, and $E_s$ is the symbol energy. The details on the variance calculation will be provided later. After applying the MMSE filtering to ${\hat{\bf r}}^{i_{d}}_{n,m}$, the output estimate ${\hat{s}^{i_{d}}_{n,m}}$ can be directly written in terms of the time domain input variable $s^{i_{d}}_{n,m}$ as \cite{Li2024LowModulation}
\begin{align}
{\hat{s}}^{i_{d}}_{n,m} =& {\bf w}^{i_{d}}_{n,m}{\hat{\bf r}}^{i_{d}}_{n,m}\nonumber\\ &={\bf w}^{i_{d}}_{n,m}{\bf H}_{n,m}[:,l']s^{i_{d}}_{n,m}+\sum_{j=0}^{l'-1} {\bf w}^{i_{d}}_{n,m}{\bf H}_{n,m}[:,j]e_{n,(m'+j)}^{i_{d}}\nonumber\\&+ \sum_{k=0}^{l'+1} {\bf w}^{i_{d}}_{n,m}{\bf H}_{n,m}[:,k]e_{n,(m'+k)}^{i_{d}-1}+{\bf w}^{i_{d}}_{n,m}{\bf z}^{i_{d}}_{n,m} \nonumber\\=& \mu^{i_{d}}_{n,m}s^{i_{d}}_{n,m}+{\tilde {z}^{i_{d}}_{n,m}}, \label{eq38}
\end{align}
where $\mu^{i_{d}}_{n,m} = {\bf w}^{i_{d}}_{n,m}{\bf H}_{n,m}[:,l']$, ${\tilde {z}^{i_{d}}_{n,m}}$ is the MMSE-suppressed noise plus residual interference and it is assumed to be Gaussian distributed. Normalizing ${\hat{s}}^{i_{d}}_{n,m}$ in (\ref{eq38}), we then have
\begin{align}
{\hat{s}}^{i_{d}}_{n,m} = \frac{{\hat{s}}^{i_{d}}_{n,m}}{\mu^{i_{d}}_{n,m}}=s^{i_{d}}_{n,m}+\frac{{\tilde {z}^{i_{d}}_{n,m}}}{\mu^{i_{d}}_{n,m}}. \label{eq39}
\end{align}
The post-MMSE variance of ${\hat{s}}^{i_{d}}_{n,m}$ can then be simply computed by
\begin{align}
\sigma_{\tilde{z}^{i_{d}}_{n,m}}^2 = \frac{\mu^{i_{d}}_{n,m}-|\mu^{i_{d}}_{n,m}|^2}{|\mu^{i_{d}}_{n,m}|^2}E_{s}. \label{eq399}
\end{align}
After performing SIC-MMSE for the $m$-th symbol in all the other $N-1$ blocks, we then have the time domain estimated symbol vector ${\hat{\bf s}}^{i_{d}}_{m} = [{{\hat s}}^{i_{d}}_{0,m},\ldots,{{\hat s}}^{i_{d}}_{(N-1),m}]^T$
and the post-MMSE covariance matrix ${\tilde{\bf V}_{m}}$ with each diagonal element obtained from (\ref{eq399}), i.e., ${\tilde{\bf V}_{m}}=$ {diag} $\left\{\sigma_{\tilde{z}^{i_{d}}_{0,m}}^2,\ldots,\sigma_{\tilde{z}^{i_{d}}_{(N-1),m}}^2\right\}$. Then, ${\hat{\bf s}}^{i_{d}}_{m+1}$ will be transformed into the DD domain and the DD domain detection can be carried out in a symbol-by-symbol fashion based on maximum likelihood. Then, we can obtain the \emph{a posteriori} mean a.k.a. the soft estimate signal vector ${\tilde{{\bf x}}}_{m}$ and the \emph{a posteriori} covariance matrix ${\bf V}_{m}$. Then, ${\tilde{\bf x}}_{m}$ is transformed back to the time domain ${\tilde{\bf s}}^{i_{d}}_{m}$ via $N$-point inverse discrete Fourier transform. The time domain covariance matrix is calculated by \begin{align}
{{\bf V}^{i_{d}}_m}= {{\bf F}_{N}^{\dag}}{\bf V}_{m}{\bf F}_{N}.\label{eq355}
\end{align}
where ${\bf F}_{N}$ and ${\bf F}_{N}^{\dag}$ are the $N$-point normalized DFT matrix and IDFT matrix, respectively. Note that the non-diagonal elements in ${{\bf V}^{i_{d}}_m}$ are not of interest due to the independence assumptions for time domain variables \cite{Li2021CrossModulation}. Finally, the newly updated time domain soft estimates ${\tilde{\bf s}}^{i_{d}}_{m}$ along with their variances ${{\bf V}^{i_{d}}_m}$ will be used as the \emph{a priori} information for detecting the subsequent signal layer in (\ref{eq31}), (\ref{eq37}),  i.e., ${\bar{s}}^{i_{d}}_{n,m}={\tilde{\bf s}}^{i_{d}}_{m}[n]$ and ${\mathbb V}\{e^{i_{d}}_{n,m}\} = {\bf V}^{i_{d}}_m[n,n]$. \par Note that in \cite{Li2024LowModulation}, The detected symbol vector ${\tilde{{\bf x}}}_{m}$ and ${\bf V}_{m}$ are directly fed back to the next symbol layer detection. This instant feedback, however, poses limitations on the interaction between the detector and the decoder. If the decoder can supply additional soft decision feedback of ${\tilde{{\bf x}}}_{m}$ and ${\bf V}_{m}$ and integrate it within the detection process, the detection performance can be significantly improved. This insight drives us to develop a coding scheme that incorporates the decoder's feedback {while detecting} the next symbol layer ${{{\bf x}}}_{m+1}$. This technique will be introduced in the next subsection. Nevertheless, in the case of SCE, the decoder must wait until all the symbols ${\tilde{{\bf x}}}$ and their variances ${\bf V}$ have been updated by the detector. Only then can the decoder begin decoding the long codeword ${\bf c}$, which contains the information ${\tilde{{\bf x}}}$, and provide parallel decoding feedback (PDF) to the detector in the next iteration. More specifically, the detector log-likelihood ratio (LLR) input to the decoder can be represented as ${\bf L}^{\text{in}}\in \mathbb {C}^{(M-L)\times N\times \log_{2}\mathcal{|Q|}}$ which is a vector that contains all the \emph{a posteriori} information from the detector. The decoder is then decoding the long codeword ${\bf c}$ based on ${\bf L}^{\text{in}}$. After decoding, the decoder output $\tilde{{\bf L}}^{\text{out}}$ is altogether in parallel (thus the name `PDF') fed back into the detector as the \emph{a priori} information during the following detection iteration, if any, or output as the final decisions if no turbo iterations are followed. In our simulations, we show that our proposed PCE with successive decoding feedback (PCE-SDF) can achieve almost up to 1dB gain compared to a single codeword encoding with parallel decoding feedback (SCE-PDF). This illustrates an intriguing phenomenon in iterative detection and decoding: multiple shorter codewords can, under certain circumstances (such as more frequent interaction with the detector), even outperform a longer codeword.
\subsection{Parallel Encoding for ODDM}\label{subsec:PEC}
In this subsection, we propose a transmission scheme featuring a parallel channel encoding (PCE) that can be particularly beneficial for the two-dimensional ODDM modulation. The block diagram of the proposed transmitter is shown in Fig. \ref{f:fig6}. Specifically, let $\mathcal{C}$ be an $(n_{c},k_{c})$ channel codes with codeword length $n_{c}$ and information length $k_{c}$. A total length of $(M-L)k_{c}$ binary information sequence ${\bf u}$ is first converted into parallel information sequences. Each parallel information sequence block ${\bf u}_{m}, m \in \{0,\ldots,M-L-1\}$, serves as the source bits for the $m$-th delay/row symbol vector ${\bf x}_{m}$ in the DD domain. Then, each ${{\bf u}_{m}}$ is encoded into a length $n_{c}$ codeword ${\mathbf c}_{m}$ of $\mathcal{C}$. The codeword ${\bf c}_{m}$ is then interleaved and becomes ${\tilde{{\bf c}}_{m}}$, which is then mapped to the $\mathcal{Q}$-QAM constellations and is formed as the $m$-th delay/row symbol vector ${\bf x}_{m}$ of length $N$ in the DD domain. Here, we have $N = n_{c}/\log_{2}{|\mathcal{Q}|}$. Once all the DD domain discrete symbols are organized, the previous ODDM modulation will be applied. This process transforms these discrete symbols into a continuous ODDM waveform. The parallel encoding structure is pivotal in our proposed joint detection and decoding algorithm, as elaborated in the next subsection. \par
\subsection{Joint Detection and Decoding for Parallel Encoded ODDM}

\begin{figure*}[!t]
\begin{minipage}[t]{0.49\linewidth}
  \centering
  \includegraphics[width=\linewidth]{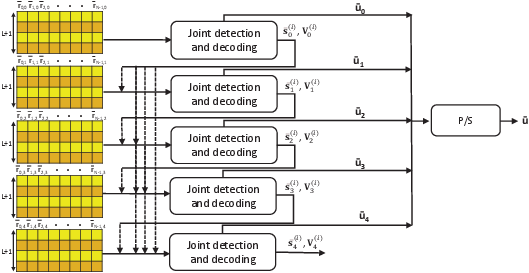}\\
 \caption{Successive decoding feedback (SDF) joint detection and decoding receiver for ZP-ODDM ($M=5$)}
 \label{f:fig4}
  \end{minipage}
  \hfill
  \begin{minipage}[t]{0.49\linewidth}
  \centering
  \centerline{\includegraphics[width=\linewidth]{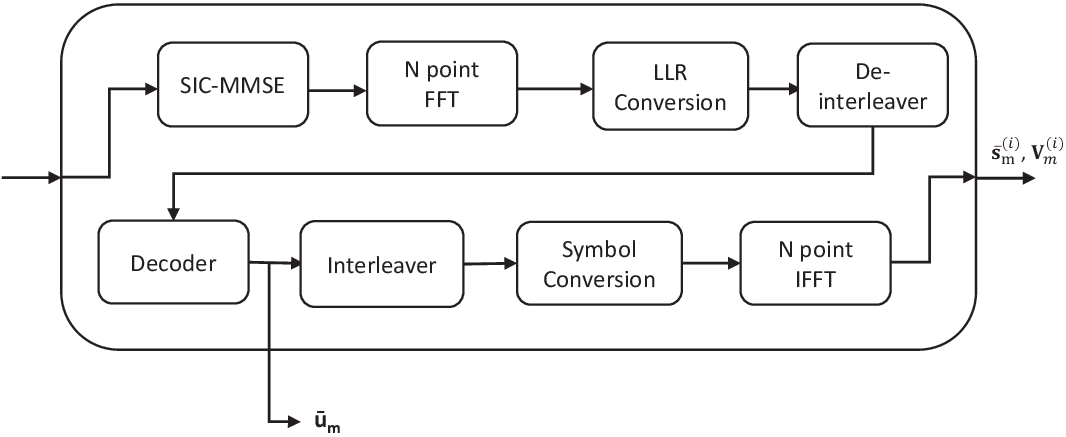}}
\caption{Joint detection and decoding for each codeword}
 \label{f:fig5}
\end{minipage}%
\end{figure*}

As introduced in Sec. \ref{subsec:MMSE}, the soft decision is performed in the DD domain, and each symbol layer ${\hat{s}}^{i_{d}}_{n,m}$ detected in the $n$-th block is held together as a vector ${\hat{\bf s}^{i_{d}}_{m}}$ and transformed into DD domain via an $N$-point DFT. This soft information is then updated and used as \emph{a priori} for the $(m+1)$-th layer detection in the time domain. Therefore, the statistical feedback from DD domain soft decisions ${\tilde{{\bf x}}}_{m}$ is crucial for the subsequent layer detection. If the DD domain information feedback is not accurate, error propagation appears and consequently, the accumulated errors will negatively impact all the subsequent layers' detection. To reduce such error propagation effect, we propose a joint detection and decoding that features instant feedback from the decoder, utilizing the unique PCE structure introduced in Sec. \ref{subsec:PEC}. 

On the receiver side, the time domain received signals ${{\bf r}_{n}}$ are first processed by SIC-MMSE as introduced in Sec. \ref{subsec:MMSE}. The channel diversity renders each transmit symbol $s_{n,m}$ to spread into a vector of received symbols $\bar{{\bf r}}_{n,m} \in \mathbb {C}^{(L+1)\times 1}$, which is defined in (\ref{eq30}). Given $\bar{{\bf r}}_{n,m}$ and the perfect channel state information (CSI) ${\bf H}_{n,m}$, when a symbol layer $s_{n,m}$ is detected by SIC-MMSE in the time domain, it will be transformed into DD domain along with the rest of $N-1$ symbols from the other parallel blocks and their soft decisions (mean and variances) are then updated. Due to the unique PCE structure, the detected symbol vector ${\tilde{\bf x}}_{m}$ also comes from a codeword ${{\bf c}_{m}}$, thus they can be immediately decoded and the decoded soft decision ${\hat{{{\bf x}}}}_{m}$ (see (\ref{eq54})) and their variances can be utilized as the new \emph{a priori} for the detection of the $(m+1)$-th layer symbol $s_{n,(m+1)}$. More specifically, the LLR of the $p$-th bit in symbol ${\tilde{{x}}}_{m}[n]$ detected by SIC-MMSE in ${\tilde{\bf x}}_{m}$ is calculated as
\begin{align}
L_{m,n}[p] &= \log_2\frac{\text {Pr}(b_p=0|\hat{y}_{m}[n])}{\text {Pr}(b_p=1|\hat{y}_{m}[n])} \nonumber\\&= \log_2\frac{\sum_{q_0\in \mathcal{Q}^{p}_0}\text {exp}(-|\hat{y}_{m}[n]-q_0|^2/\tilde{{\bf V}}_{m}[n,n])}{\sum_{q_1\in \mathcal{Q}^{p}_1}\text {exp}(-|\hat{y}_{m}[n]-q_1|^2/\tilde{{\bf V}}_{m}[n,n])},\label{eq51}
\end{align}
where $b_p$ represents the bit value of the $p$-th bit, $\mathcal{Q}^{p}_0$ is the symbol set that has the $p$-th bit as zero and $\mathcal{Q}^{p}_1$ is the symbol set that has the $p$-th bit as one, $\hat{y}_{m}[n]$ is the MMSE filtered $n$-th symbol at the $m$-th layer in DD domain. Using (\ref{eq51}) for $n=0,\ldots,N-1$ and $p =1,\ldots,\log_2|\mathcal{Q}|$, we obtain the LLR vector ${\bf L}^{\text{in}}_{m}$ for the entire $m$-th layer $N$ symbols. After de-interleaving, ${\bf L}^{\text{in}}_{m}$ is used as the decoder's \emph{a priori} information. The decoder will then perform the decoding algorithm, and after certain decoding iterations, the decoder generates new output LLRs ${\bf L}^{\text{out}}_{m}$, which will then be first converted to symbols and used for detecting the next layer's symbol ${\tilde{\bf x}}_{m+1}$. Note that the \emph{a posteriori} ${\bf L}^{\text{out}}_{m}$ is fed back to the detector instead of extrinsic information. The reason is that the higher posterior values cancel out more interference for high interference ODDM channel, {which} was also revealed in \cite{Witzke2002IterativeDetectors}. Nevertheless, we refer to such decoding feedback scheme as successive decoding feedback (SDF), which indicates that the decoder sequentially decodes each detected layer ${\tilde{\bf x}}_{m}$. With the decoder output acquired, ${\bf L}^{\text{out}}_{m}$ is first re-interleaved becoming $\tilde{{\bf L}}^{\text{out}}_{m}$ and then converted to soft symbol estimates by
\begin{align}
  {\hat{{x}}}_{m}[n] = \sum_{q\in \mathcal{Q}}q\times \frac{1}{2^{\text{log}_{2}|\mathcal{Q}|}} \prod_{p=1}^{\text{log}_{2}|\mathcal{Q}|}\left(1+{\tilde d}_{p}{\text {tanh}}\left(\frac{\tilde{L}_{m,n}^{\text{out}}[p]}{2}\right)\right),\label{eq54}
\end{align}
where the subscript $p$ indicates the association to the $p$-th bit of constellation symbol $q\in \mathcal{Q}$ and ${\tilde d}_{p} \in \{1,-1\}$. The variance can then be computed by
\begin{align}
\sigma^2_{\hat{x}_{m}[n]} =& \sum_{q\in \mathcal{Q}}|q|^2\times \frac{1}{2^{\text{log}_{2}|\mathcal{Q}|}} \prod_{p=1}^{\text{log}_{2}|\mathcal{Q}|}\left((1+{\tilde d}_{p}{\text {tanh}}\left(\frac{\tilde{L}^{\text{out}}_{m,n}[p]}{2}\right)\right)\nonumber \\ &-{\hat{{x}}}_{m}[n]|^2.\label{eq55}
\end{align}
These $N$ new soft estimates ${\bf x}_{m}$ and variances $\sigma^2_{x_{m}}$ for the $m$-th layer symbol will be first converted to the time domain and used as \emph{a priori} information for detecting the subsequent symbol layer ${\tilde{\bf x}}_{m+1}$ during interference cancellation in (\ref{eq31}) and MMSE weighs calculation in (\ref{eq37}). After all the symbols have been processed, this will complete the first round of turbo iteration for the PC-SDF scheme. Fig. \ref{f:fig4} shows the proposed SDF block diagram during the $i_{d}$-th iteration. Fig. \ref{f:fig5} shows how the $m$-th layer symbol is processed by the detector and decoder during the $i_{d}$-th turbo iteration. Another turbo iteration will be initiated if the maximum allowed iterations have not been reached. When the maximum turbo iteration is reached, the soft outputs from the decoder in the last iteration are mapped to the hard decision $\bar{{\bf u}}_{m}\in \mathbb {F}^{k_{c}}_{2}$ and all the $M-L$ decoded codewords will be concatenated through parallel to serial conversion.\par

\begin{remark}
{Without loss of generality, the PCE transmission structure plus the detection and decoding strategy proposed for ZP-ODDM is also applicable to CP-ODDM . In CP based systems, interferences may emerge among different blocks, rendering all sub-channel sizes to be $(2L+1)\times(2L+1)$, and this channel condition aligns with the scenario encountered by signal layers when the index of targeted signal layer $m$ exceeds ‘$L$’ in ZP model already. Due to the nature of layer-by-layer detection and interference cancellation in SIC-MMSE, PCE-SDF presented in this paper remains applicable and effective for other alternative systems.}
\end{remark}

\section{Performance Analysis}\label{sec:analysis}
In this section, we conduct an analysis of the SCE-PDF and PCE-SDF schemes utilizing density evolution (DE) with Gaussian approximation. We model the detector output as a symmetric Gaussian mixture and employ DE to track the probability density function (PDF). Our analysis closely aligns with the simulation results, indicating that it accurately characterizes the message exchanged between the detector and decoder.\par
\subsection{SCE-PDF Analysis}
\begin{figure}[!t]
\includegraphics[width=\linewidth]{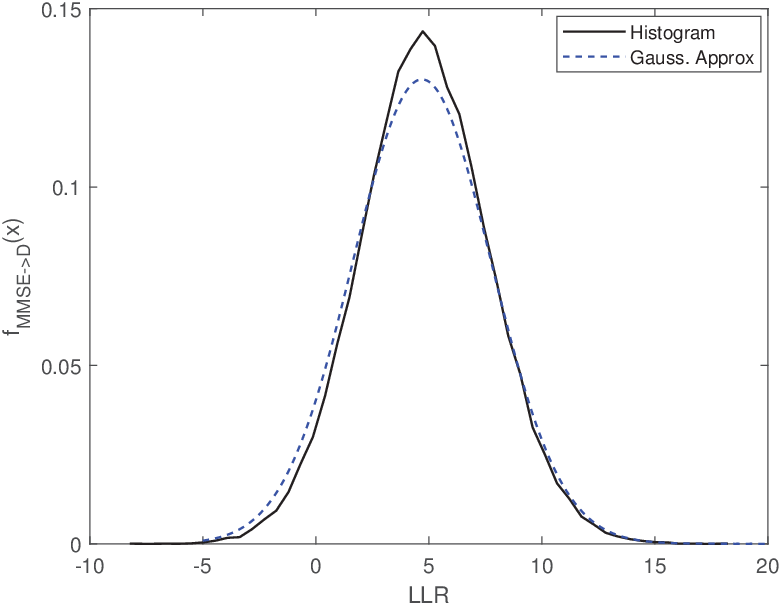}
\caption{Histogram for the SIC-MMSE detector output for SCE ODDM with $M=132$ and $N=648$, and the symmetric Gaussian approximations using EM.}
\label{f:fig11187}
\end{figure}

We first analyze the performance of the benchmark SCE scheme. We have demonstrated that the output of SIC-MMSE filtering can be expressed as (\ref{eq38}). Due to the unitary transformation between the time domain and the DD domain, the DD domain observation $\hat{y}_{m}[n]$ can be approximated as a Gaussian distributed random variable, i.e.
\begin{align}
p(\hat{y}_{m}[n]) \approx {\cal{CN}}({{{x}}}_{m}[n],\sigma^2_{{x}_{m}[n]}),\label{eq44}
\end{align}
where $\sigma^2_{{x}_{m}[n]} = \frac{\mu_{n,m}-|\mu_{n,m}|^2}{|\mu_{n,m}|^2}E_{s}$ which is the same as the time domain $\sigma_{\tilde{z}^{i_{d}}_{n,m}}^2$, and $\mu_{n,m} = {\bf w}_{n,m}{\bf H}_{n,m}[:,l']$, see (\ref{eq38}). Take the example of BPSK, we then have the LLR of each bit ${x}_{m}[n]$ calculated as
\begin{align}
L({x}_{m}[n]) &= \text {log}\frac{\text {Pr}(\hat{y}_{m}[n]|{x}_{m}[n]=+1)}{\text {Pr}(\hat{y}_{m}[n]|{x}_{m}[n]=-1)}\\
&= -\frac{|\hat{y}_{m}[n]-1|^2}{\sigma^2_{{x}_{m}[n]}}+\frac{|\hat{y}_{m}[n]+1|^2}{\sigma^2_{{x}_{m}[n]}}\nonumber\\&=\frac{4\mu_{n,m}\Re\{\hat{y}_{m}[n]\}}{1-\mu_{n,m}}.\label{eq45}
\end{align}
The real part of $\Re\{\hat{y}_{m}[n]\}$ follows the Gaussian distribution
\begin{align}
\Re\{\hat{y}_{m}[n]\}\sim{\cal{N}}({x}_{m}[n],\frac{\sigma^2_{{x}_{m}[n]}}{2}).\label{eq46}
\end{align}
Hence, the pdf $f_{\text{LLR}}$ of each bit's LLR also adheres to a Gaussian distribution with mean and variance as follows:
\begin{align}
E\{L({x}_{m}[n])\} &= \biggl(\frac{4\mu_{n,m}}{1-\mu_{n,m}}\biggl)E\{\Re\{\hat{y}_{m}[n]\}\} \nonumber\\ &= \frac{4\mu_{n,m}{x}_{m}[n]}{1-\mu_{n,m}},\\\text{Var}\{L({x}_{m}[n])\} &= \biggl(\frac{4\mu_{n,m}}{1-\mu_{n,m}}\biggl)^2\text{Var}\{\Re\{\hat{y}_{m}[n]\}\} \nonumber\\ &= \frac{8\mu_{n,m}}{1-\mu_{n,m}}.
  \label{eq47}
\end{align}
Thus, the pdf $f_{\text{LLR}}$ of each bit's LLR from SIC-MMSE output can be written as $L({x}_{m}[n])\sim {\cal{N}}\biggl(\frac{4\mu_{n,m}{x}_{m}[n]}{1-\mu_{n,m}},\frac{8\mu_{n,m}}{1-\mu_{n,m}}\biggl)$. Denote $f_{\text{MMSE}\rightarrow \text{D}}$ as the pdf of SIC-MMSE detector output LLR, $f_{\text{MMSE}\rightarrow \text{D}}$ is then obtained by adding each individual Gaussian pdf $f_{\text{LLR}}$. To approximate $f_{\text{MMSE}\rightarrow \text{D}}$ for the sequent analysis, a mixture of symmetric Gaussian pdf.  In this case, the output from the detector using a single pdf expression can be written as \cite{Narayanan2005EstimatingEqualization}
\begin{align}
f_{\text{MMSE}\rightarrow \text{D}} \triangleq \sum_{j=1}^{J}\pi_{j}{\cal{N}}(\mu_{j},2\mu_{j}),
  \label{eq48}
\end{align}
where $J$ is the number of mixtures and we are using a finite $J$ to have a close approximation. ${\cal{N}}(\mu_{j},2\mu_{j})$ is the symmetric Gaussian pdf and $\pi_{j}$ is the weight of each term. Based on the collections of the LLRs $\Xi \triangleq \{L({x}_{m}[n]),m = 0,\ldots, M-L-1, n = 0,\ldots, N-1\}\in \mathbb {R}^{(M-L)\times N\times \log_{2}|\mathcal{Q}|}$, the parameters $\{\pi_{j},\mu_{j},j=1,\ldots, J\}$ can then be estimated using the expectation-maximization (EM) algorithm \cite{Lu2004PerformanceSystems}. In Fig. \ref{f:fig11187}, we show that the proposed Gaussian mixture approximation for the detector output closely matches its histogram. \par
With the output from the detector, we can now trace the pdf of the extrinsic information within the decoder using density evolution (DE) \cite{Chung2001AnalysisApproximation}. To simplify the process, we make the assumption that an all-zeros codeword is transmitted. However, as the coded ODDM modulation channel is not symmetric, we apply the concept of an independent and identically distributed (i.i.d.) channel adapter \cite{Hou2003Capacity-ApproachingCodes}. By maintaining the assumption of an all-zeros codeword and incorporating the channel adapter, we can iterate between the bit node and check node following the steps below. Considering an LDPC code with degree profiles $\lambda(x) = \sum\limits_{i=2}^{d_{l,\max}}\lambda_{i}x^{i-1}$, $\rho(x) = \sum\limits_{i=2}^{d_{r,\max}}\rho_{i}x^{i-1}$, where $\lambda_{i}$ and $\rho_{i}$ are the fractions of edges belonging to degree-$i$ bit and check nodes, respectively. The steps of performing DE are as follows: \par
\begin{itemize}
\item {\emph {At a bit node of degree i}}: Assuming that the incoming messages from the check node $c$ to bit node $b$ are Gaussian with mean $m_{b\leftarrow c}^{i_{c}-1,i_{d}}$, where $i_{c}$ and $i_{d}$ are the LDPC decoding and detection iterations, respectively. The pdf of the extrinsic message at the bit node of degree $i$ is \cite{Lu2004PerformanceSystems}
\begin{align}
  f_{b\rightarrow c, i}^{i_{c}, i_{d}} = \sum\limits_{j=1}^{J}\pi_{j}{\cal N} \left(\mu_{j}+(i-1)m_{b\leftarrow c}^{i_{c}-1, i_{d}},\right. \nonumber\\ 2\left. \left[\mu_{j}+(i-1)m_{b\leftarrow c}^{i_{c}-1, i_{d}}\right]\right).\label{eq49}
\end{align}
Since a fraction $\lambda_{i}$ of the edges are connected to bit nodes of degree $i$, the pdf of the extrinsic message passed from the bit nodes to the check nodes along an edge is
\begin{align}
  f_{b\rightarrow c}^{i_{c}, i_{d}}=&\sum\limits_{j=1}^{J}\sum\limits_{i=2}^{d_{l,\max}}\pi_{j}\lambda_{i}{\cal N}\left(\mu_{j}+(i-1)m_{b\leftarrow c}^{i_{c}-1, i_{d}} , 2 [\mu_{j}\right.\nonumber\\
  &+\left.(i-1)m_{b\leftarrow c}^{i_{c}-1, i_{d}}]\right).\label{eq410}
\end{align}\par
\item {\emph {At a check node of degree j}}: Let $\psi(x)\triangleq E\{\tanh(1/2{\cal N}(x,2x))\}$. Assume that the outgoing message has a Gaussian density. Then, the mean of the outgoing messages at a check node of degree $j$ is
\begin{align}
  m_{b\leftarrow c, j}^{i_{c}, i_{d}}=\psi^{-1}\left[\left(\sum_{l=1}^{J}\sum\limits_{i=2}^{d_{l,\max}} \pi_{l}\lambda_{i}\psi \left(m_{b\rightarrow c, i}^{i_{c}, i_{d}}\right)\right)^{j-1}\right].\label{eq411}
\end{align}
Since a fraction $\rho_{j}$ of the edges are connected to check nodes of degree $j$, the mean of the extrinsic message passed from the check node to the bit node (averaged over the degrees of the check nodes) is
\begin{align}
 m_{b\leftarrow c}^{i_{c}, i_{d}}=\sum\limits_{j=2}^{d_{r,\max}}\ \rho_{j}m_{b\leftarrow c, j}^{i_{c}, i_{d}}.\label{eq412}
\end{align}\par
\item \emph {Message passed back to the detector}: At a bit node of degree $i$, the mean of the extrinsic message passed back to the detector is
\begin{align}
  m^{i_{d}}_{\text{MMSE}\leftarrow \text{D}}(i) = im^{i_{c}-1,i_{d}}_{b\leftarrow c}.
\end{align}
The averaged mean passed back to the detector is then
\begin{align}
m^{i_{d}}_{\text{MMSE}\leftarrow \text{D}}=\sum\limits_{i=2}^{d_{l,\max}}{\tilde\lambda}_{i}m^{i_{d}}_{\text{MMSE}\leftarrow \text{D}}(i),
\end{align}
where ${\tilde\lambda}_{i}$ is the fraction of variable nodes with degree $i$. Once the turbo iteration reaches the maximum, i.e. $i_{d}=i_{d,\max}$, we then calculate the error probability $P_{b}$ given the intrinsic information $m^{i_{d,\max},\text{intr}}_{\text{MMSE}\leftarrow \text{D}}$
\begin{align}
m^{i_{d,\max},\text{intr}}_{\text{MMSE}\leftarrow \text{D}}=m^{i_{d,\max}}_{\text{MMSE}\leftarrow \text{D}}+\sum\limits_{j=1}^{J}\pi_{j}\mu_{j},\label{eq414}
\end{align}
where $\sum\limits_{j=1}^{J}\pi_{j}\mu_{j}$ is the message from the channel. Since the transmitted bits are all zeros, the error probability $P_{b}$ is then estimated as
\begin{align}
P_{b} = \frac{1}{1+e^{m^{i_{d,\max},\text{intr}}_{\text{MMSE}\leftarrow \text{D}}}}.\label{eq415}
\end{align}
\end{itemize}

\subsection{PCE-SDF Analysis}The process of deriving an analytical error probability $P_{b}$ for the PCE-SDF scheme is akin to the analysis for the SCE-PDF scheme. However, different from the SCE, the message here passed back to the detector is the $m$-th row's decoding information $m^{i_{d}}_{\text{MMSE}\leftarrow \text{D},m}$, which will be utilized as the \emph{a priori} information for detecting the subsequent row. Due to the PC-SDF scheme comprising multiple short codewords, we cannot overlook the finite-length property of the code, where the channel behavior experienced by each codeword differs. Thus, we employ a modified version of DE analysis. As previously discussed, the SIC-MMSE output for each layer can be approximated as Gaussian distributed
\begin{align}
{\hat{\bold{x}}}^{i_{d}}_{m} = {\bold x}^{i_{d}}_{m}+{\tilde {\bold{n}}^{i_{d}}_{m}}, \label{eq416}
\end{align}
where ${\bold{x}}^{i_{d}}_{m}$ is the time domain transmitted symbols and the LLR for each bit originates from the short length codeword ${\mathbf c}_{m}$, which usually cannot be well analyzed with DE. ${\tilde {\bold{n}}^{i_{d}}_{m}}$ is the Gaussian approximated noise plus interference with averaged variance $\sigma^2_{x_m}$.
If the length of the codeword $n_{c}$ approaches infinity, i.e. $n_{c} \rightarrow \infty$, the observed error rate of each codeword after transmission and before decoding stabilizes at $P_{\text{obs}} = p_{0} = Q(1/\sigma_{x_m})$, where $Q(.)$ denotes the $Q$ function. If the length of the codeword $n_{c}$ is finite, $P_{\text{obs}}$ varies from one codeword to another, thus can be represented by a random variable distributed around $p_{0}$, i.e. $P_{\text{obs}}\sim {\cal{N}}(p_{0},p_{0}(1-p_{0})/n_{c})$ \cite{Yazdani2009WaterfallChannels}. The block error probability is thus the probability when $P_{\text{obs}} > p_{th}$, that is
\begin{align}
P_{B} = Q\left(\frac{p_{th}-\mu_{P_{\text{obs}}}}{\sigma_{P_{\text{obs}}}}\right), \label{eq417}
\end{align}
where $\mu_{P_{\text{obs}}}$ is the mean of $P_{\text{obs}}$, and
$\sigma^2_{P_{\text{obs}}} = p_{0}(1-p_{0})/n_{c}$. $p_{th}$ is the channel error probability at the decoding threshold under AWGN, which can be found using DE. Then, we derive the fraction of bits in error $\alpha$ for each codeword by setting the channel parameters worse than (but close to) the threshold. $\alpha$ can be calculated as
\begin{align}
\alpha = \sum\limits_{i=2}^{d_{l,\max}}{\tilde\lambda}_{i}Q\Biggl(\sqrt{\frac{m_{\mu_{0}}+im^{i_{c},i_{d}}_{b\leftarrow c}}{2}}\Biggl), 
\end{align}
where $m_{\mu_{0}} = \frac{2}{\sigma_{x_{m}}^2}$ is the mean of the channel messages. Finally, the average estimated bit error probability from the $m$-th row's codeword is $P_{b,m} = \alpha P_{B}$. \par
Following the previous finite length analysis for PCE, we can obtain the average bit error probability for $P_{b,m}$, $m\in\{0,\ldots,M-L-1\}$. The averaged bit error probability for each ODDM frame is thus
\begin{align}
P_{b,\text{avg}} = \frac{1}{M-L}\sum\limits_{m=0}^{M-L-1}P_{b,m}. 
\end{align}

In Sec. \ref{sec:sim}, we demonstrate that both proposed analytical methods for SCE and PCE can accurately predict the simulation error performance. Additionally, the finite-length analysis adds further accuracy specifically for PCE.

\subsection{Complexity Analysis}

\begin{figure}[!t]
\includegraphics[width=\linewidth]{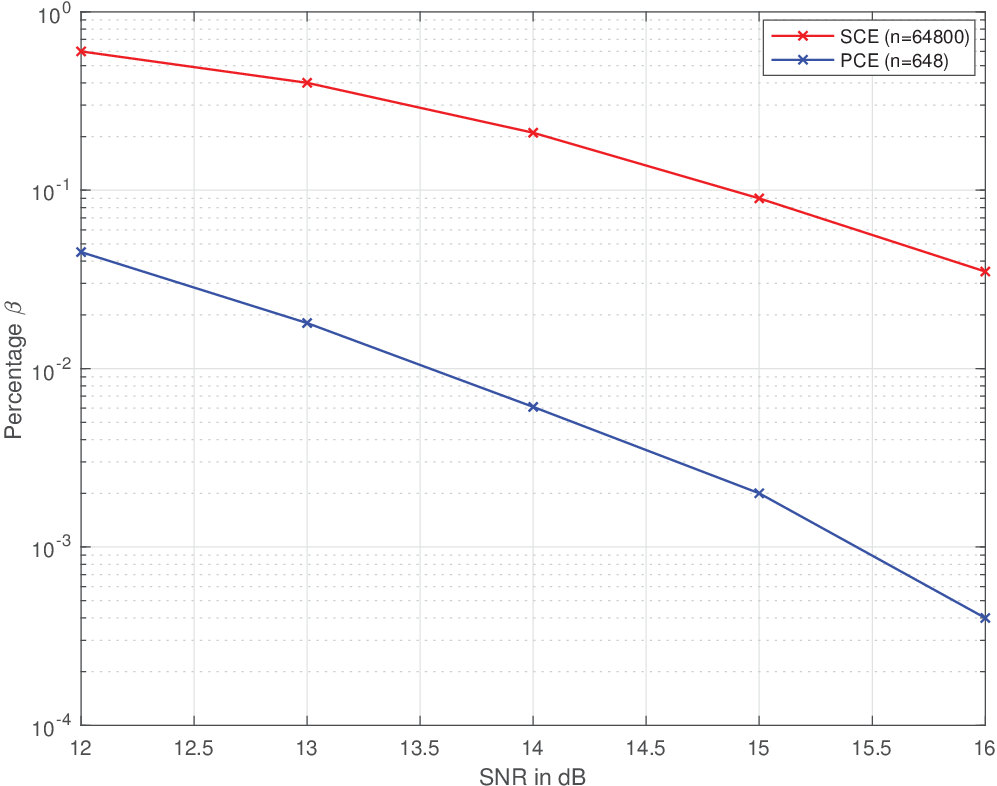}
\caption{Percentage of codeword that do not satisfy parity check after the first turbo iteration. The system is rate-$\frac{1}{2}$ LDPC encoded BPSK ODDM with $M=132$ and $N=648$. }
\label{f:fig111}
\end{figure}
\begin{figure}[!t]
  \centering
  \includegraphics[width=\linewidth]{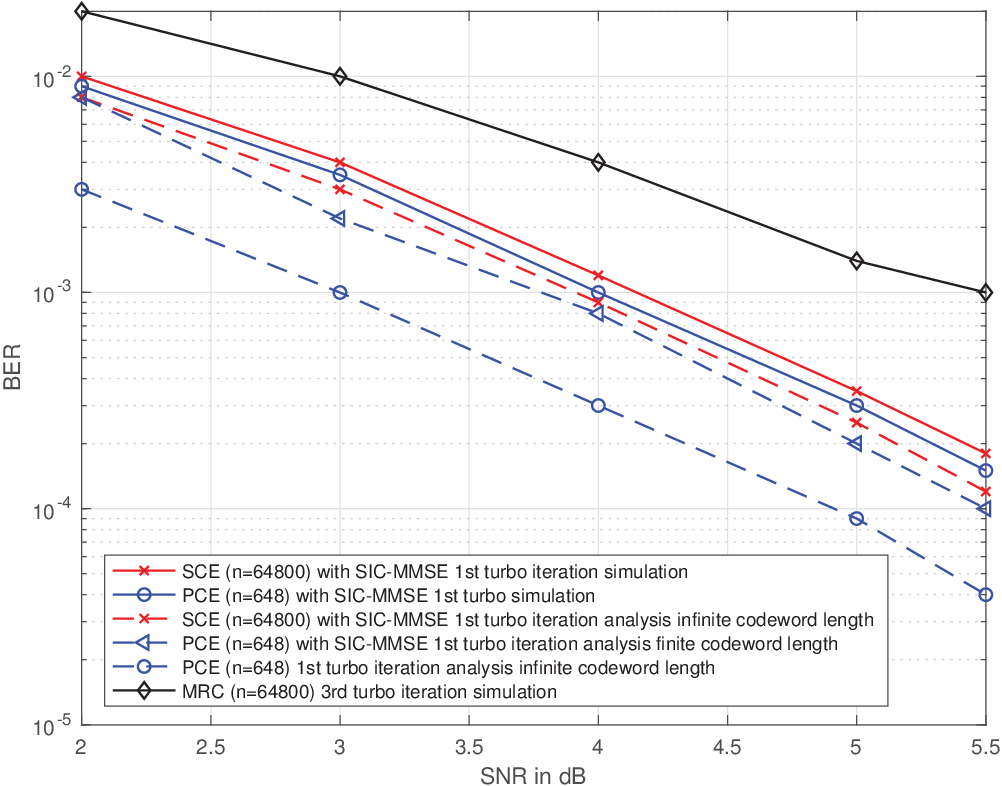}\\
 \caption{Performance analysis using DE, and simulation results for rate-$\frac{1}{2}$ LDPC encoded BPSK ODDM with $M=132$ and $N=648$.}
\label{f:fig11}
  \end{figure}
  
The computational complexity for the interference cancellation in (\ref{eq31}) can be reduced by storing $\Delta {\bf r} = {\bf r}-\sum\limits_{j=1}^{l'+1}{\bf H}_{n,m}[:,j]{\tilde s}_{j}^{(i)}$ as an interference-reduced variable so that ${\bf r}$ does not have to cancel interference repeatedly each time for new target symbols. Let $L=l_{max}$ such that the number of zero padding equal to the maximum number of delay taps. Consequently, this only requires $\mathcal{O}((M-l_{max})Nl_{max})$ complex multiplications.  Furthermore, the calculation of MMSE weights entails an overall complexity of $\mathcal{O}((M-l_{max})Nl_{max}^3)$. Besides MMSE weight computations, determining the MMSE output ${\hat{s}}^{(i)}_{n,m}$ and variance $\sigma_{\tilde{z}^{(i)}_{n,m}}^2$ for all symbols requires another $\mathcal{O}(2(M-l_{max})Nl_{max})$ complex multiplications. The transformation between the time domain and the DD domain needs two $N$-FFT operations which have a complexity of $\mathcal{O}(2(M-l_{max})N\log_{2}N)$. \par At the decoder, without considering the convergence behavior from the previous turbo iteration, the decoding complexity for the PCE short codeword scheme is $\mathcal{O}{({IN}})$ per codeword and overall complexity of $\mathcal{O}({I(M-l_{max})N})$, assuming BP decoding is used with $I$ being the number of decoding iteration. In contrast, the overall decoding complexity for SCE is $\mathcal{O}({I(M-l_{max})N})$ per turbo iteration due to its single $(M-l_{max})N$ length codeword. However, when considering the convergence behavior, the PCE scheme has lower detection and decoding complexity than the SCE scheme. Specifically, from the second turbo iteration and after, the codeword that satisfies the parity check from the last turbo iteration does not need to be \emph{repeatedly} decoded and its corresponding symbols do not need to be detected. Only the codewords and symbols from the last iteration that do not satisfy the parity check will be sent into the decoder and detector, respectively, for further processing. However, this is not possible in the SCE scheme. Therefore, the row-by-row encoded PCE offers more flexibility for each codeword to converge faster than SCE and thus significantly reduces the overall complexity. 

Denote $\beta_{i_{d},\text{s}}$ and $\beta_{i_{d},\text{p}}$ as the average percentage of the codewords that do not satisfy the parity check from the $i_{d}$-th iteration for SCE and PCE, respectively. The overall complexity for SCE and PCE, when taking the decoding convergence into account, is thus dominated by $\mathcal{O}(\beta_{i_{d},\text{s}}(M-l_{max})Nl_{max}^3)+\mathcal{O}({\beta_{i_{d},\text{s}}I(M-l_{max})N})$ and $\mathcal{O}(\beta_{i_{d},\text{p}}(M-l_{max})Nl_{max}^3)+\mathcal{O}({\beta_{i_{d},\text{p}}I(M-l_{max})N})$, respectively. In Fig. \ref{f:fig111}, we can see that the percentage for PCE is nearly two orders of magnitude lower than that of SCE at high SNR, meaning the overall complexity of PCE is about $\frac{1}{100}$ of that of the SCE scheme. A summary of computation complexity for both schemes is included in Table \ref{my-label}.

\begin{table*}
\centering
\caption{Detection and Decoding Complexity Comparisons for SCE and PCE}
\label{my-label}
\begin{tabular}{ | m{4.2em} | m{3.5cm}| m{3.4cm} | m{5cm}|}
  \hline
   & Decoding & Detection  & Complexity per turbo iteration \\
  \hline
  SCE & $\mathcal{O}({I(M-l_{max})N})$ & $\mathcal{O}((M-l_{max})Nl_{max}^3)$ & $\mathcal{O}({\beta_{i_{d},\text{s}}I(M-l_{max})N})+\mathcal{O}(\beta_{i_{d},\text{s}}(M-l_{max})Nl_{max}^3)$ \\
  \hline
  PCE & $\mathcal{O}({I(M-l_{max})N})$ & $\mathcal{O}((M-l_{max})Nl_{max}^3)$ & ${O}({\beta_{i_{d},\text{p}}I(M-l_{max})N})+\mathcal{O}(\beta_{i_{d},\text{p}}(M-l_{max})Nl_{max}^3)$ \\
  \hline
\end{tabular}
\end{table*}

\section{Simulations}\label{sec:sim}

 \begin{figure}[!t]
 \centering
  \includegraphics[width=\linewidth]{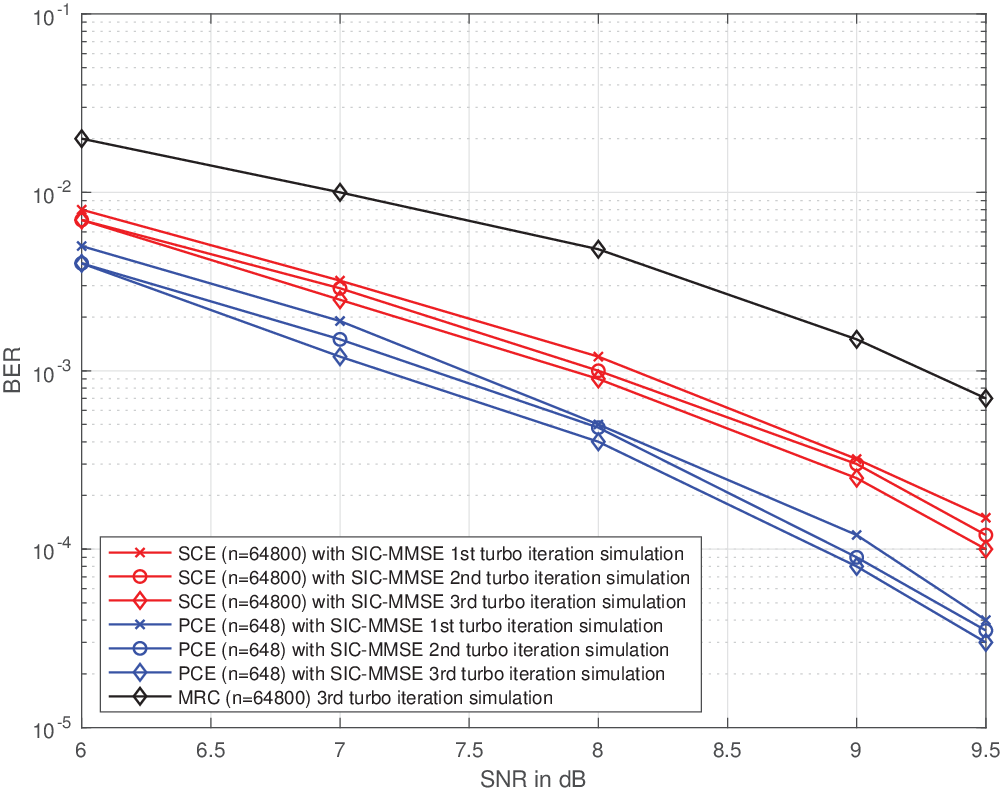}\\
 \caption{Simulation results for rate-$\frac{1}{2}$ LDPC encoded 4QAM ODDM with $M=132$ and $N=324$.}
 \label{f:fig1111}
  \end{figure}
  
\begin{figure}[!t]
  \centering
  \includegraphics[width=\linewidth]{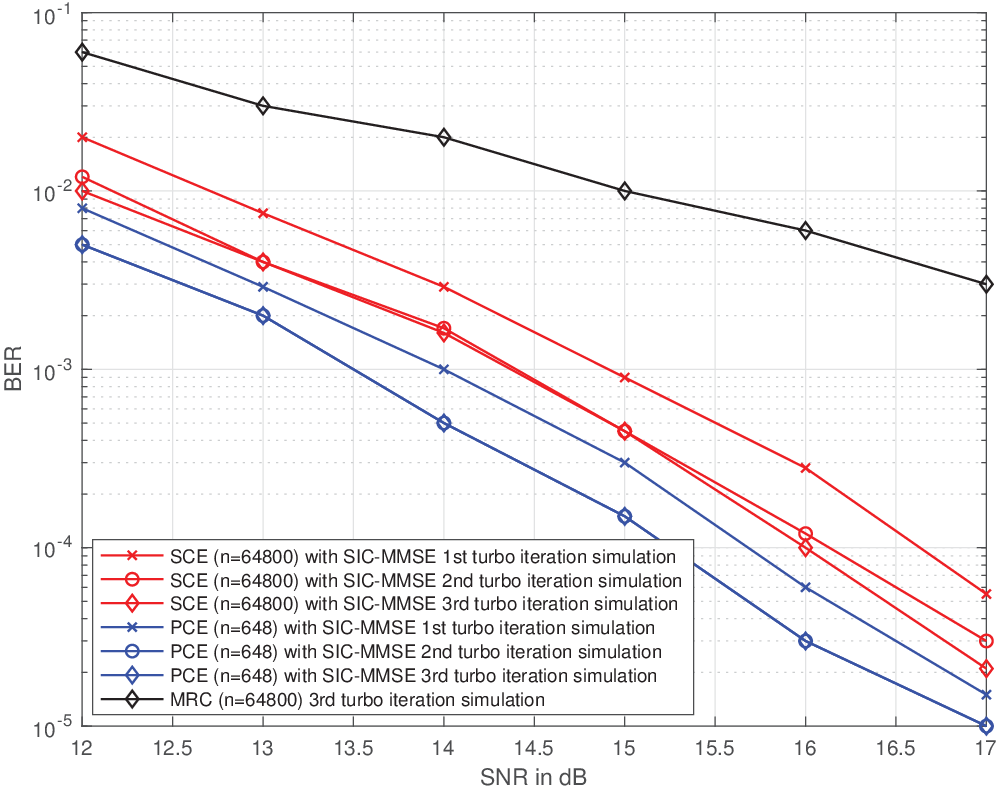}\\
 \caption{Simulation results for rate-$\frac{1}{2}$ LDPC encoded 16QAM ODDM with $M=132$ and $N=162$.}
 \label{f:fig9}
   \end{figure}

\begin{figure}[!t]
  \centering
 \includegraphics[width=\linewidth]{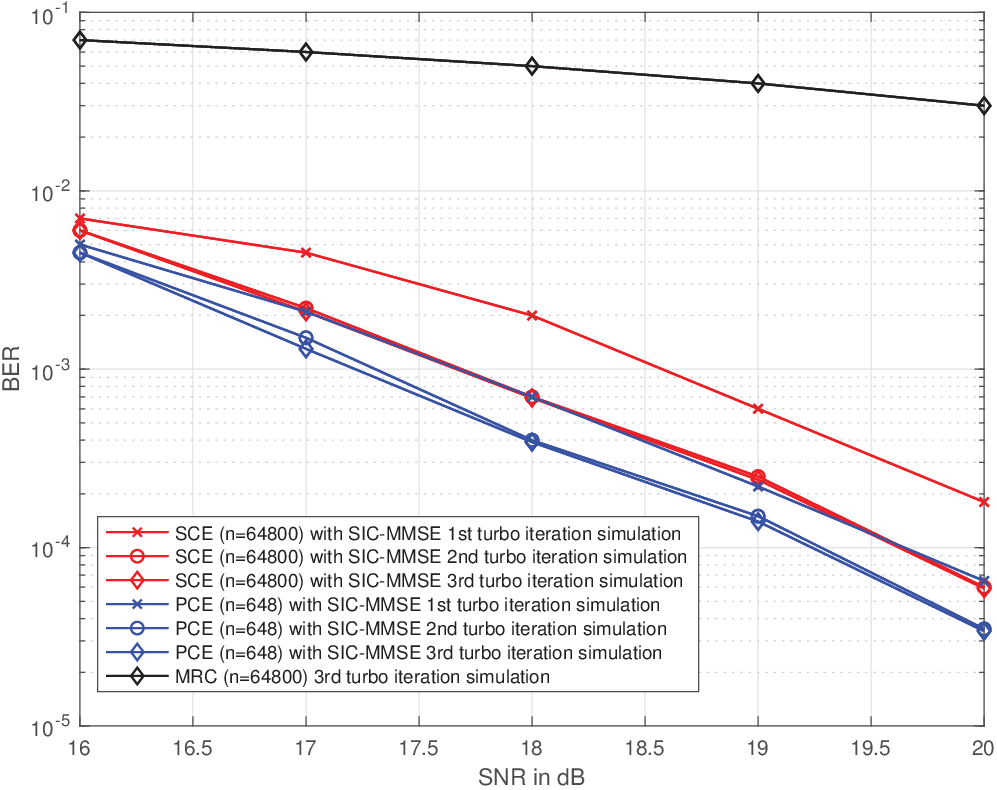}
\caption{Simulation results for rate-$\frac{3}{4}$ LDPC encoded 16QAM ODDM with $M=132$ and $N=162$.}
\label{f:fig10}
   \end{figure}

\begin{figure}[!t]
\centering
\includegraphics[width=\linewidth]{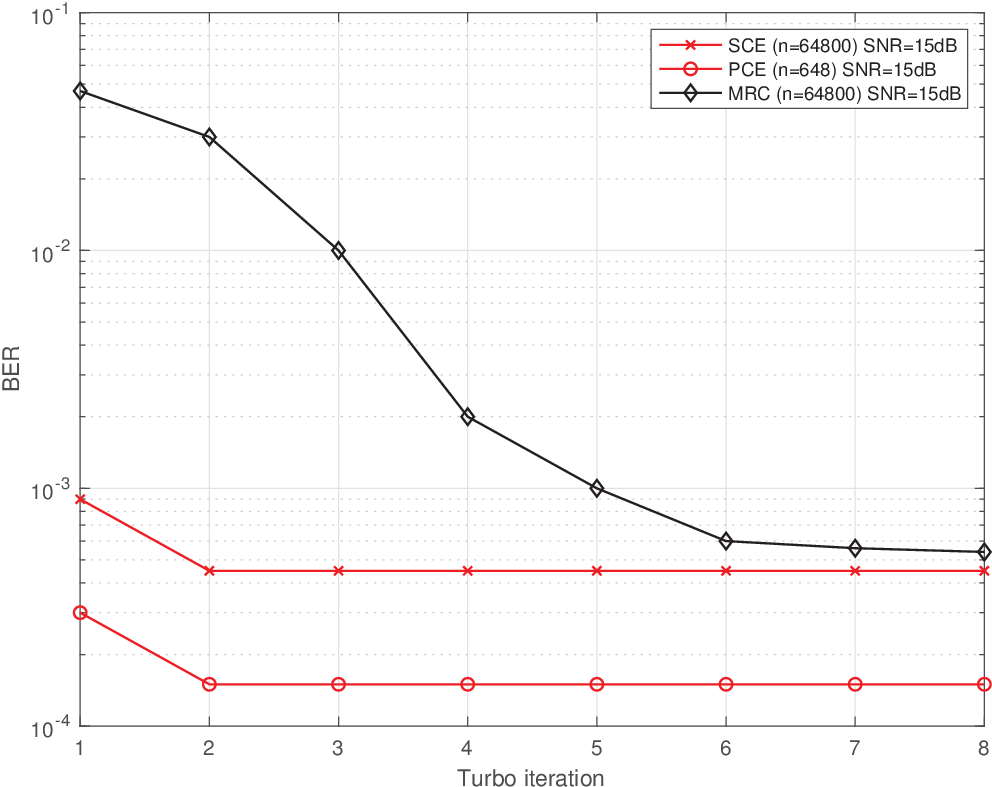}
\caption{BER simulation for 16QAM ODDM with rate-$\frac{1}{2}$ LDPC code at SNR=15dB for a range of turbo iterations.}
\label{f:fig12}
\end{figure}

In this section, we provide simulation results with different code rates for the proposed scheme for ZP-ODDM. Similar to 4G/5G standards, the subcarrier spacing $\Delta f$ is set as 15 kHz and $T=\frac{1}{\Delta f}$. The carrier frequency is $4$ GHz. The ODDM frame size is fixed with $M=132$ and $N=648, 324, 162$ for BPSK, 4QAM, 16QAM, respectively. The wireless channel is generated with nine multipaths having a maximum integer delay tap $l_{\max}=19$ and a maximum fractional Doppler tap $k_{\max}=3.79$. The Doppler taps for the $i$-th path are generated from a uniform distribution $U(-k_{\max},k_{\max})$. A perfect channel state information is assumed at the receiver. The simulation does not stop until 200 ODDM frame errors have been collected. \par For the PCE scheme, we adopt the LDPC codes used are from the IEEE 802.11 standard \cite{IEEE2020IEEESpecifications} with $n_{c}=648$, $k_{c}=486$ for 0.75 code rate and $n_{c}=648$, $k_{c}=324$ for 0.5 code rate. For comparison purposes, we include a benchmark scheme that uses a single channel encoding (SCE) to encode the whole ODDM source symbols \cite{Lin2022OrthogonalModulation}. The LDPC codes used for the benchmark scheme are from the DVB-S2 standard \cite{JTC2014ENDVB-S2} with $n=64800$, $k=48600$ for 0.75 code rate and $n=64800$, $k=32400$ for 0.5 code rate.\par For the benchmark SCE scheme, the SIC-MMSE detector first detects the whole ODDM frame ${\bf X}$ and then passes the soft estimate $\tilde{{\bf X}}$ to the decoder. Then, the decoder feedbacks the decoded soft messages to the detector using the belief-propagation (BP) algorithm. To generate a fair comparison, both schemes adopt the BP algorithm with the same decoding iterations. The number of turbo iterations is indicated in the legends of the figures. \par

Fig. \ref{f:fig11} compares the analytical BER using the methods outlined in Sec. \ref{sec:analysis} with the simulation BER for both schemes and the MRC detection. Note that the performance analysis in Sec. \ref{sec:analysis} is based on one channel realization. Here, the analytical BER for the SCE-ODDM and PCE-ODDM with BPSK is obtained by averaging over many channel realizations. It is evident that the DE analysis works effectively for the long codeword SCE scheme with codeword length of $n=64800$, but it exhibits discrepancies for the PCE scheme with codeword length $n=648$. Through adjustments to the DE method and consideration of the codeword length, the accuracy of PCE analytical results has been notably enhanced by considering the finite blocklength scaling law. It can be observed that the proposed PCE scheme achieves better BER than the benchmark SCE scheme in both analytical and simulation results even in its first iteration. The reason is that PCE-ODDM features constant information exchange between the detector and decoder, improving the overall error performance. Last but not least, both one-iteration MMSE-based detections can outperform the MRC-based detection even with three turbo iterations. Note that this agrees with the results in \cite{Huang2023PerformanceEstimation}, which shows that uncoded ODDM with SIC-MMSE achieves the better performance than that with MRC and MPA.

Fig. \ref{f:fig1111} illustrates the BER for rate-$\frac{1}{2}$ LDPC coded 4QAM modulated ODDM systems with 50 iterations of the BP decoding algorithm. It can be observed that the proposed PCE demonstrates a superior error performance across every turbo/detection iteration, compared to SCE. Notably, a single turbo iteration of our PCE scheme can surpass the error performance of three turbo iterations of the SCE scheme, thereby reducing the computational complexity needed for the PCE scheme to achieve comparable error performance to SCE. Still, both one-iteration MMSE-based detections can outperform three-iteration MRC. \par

Fig. \ref{f:fig9} shows the BER for rate-$\frac{1}{2}$ LDPC coded 16QAM modulated ODDM systems with 50 iterations of the BP decoding algorithm. It is observed that the proposed PCE-SDF still outperforms the benchmark scheme SCE-PDF across every turbo iteration, and can achieve 1 dB gain at a BER of $4\times10^{-5}$ compared to SCE-PDF. The proposed MMSE based detections significantly outperform MRC under higher-order modulation. \par

Fig. \ref{f:fig10} shows the BER for rate-$\frac{3}{4}$ LDPC coded 16QAM modulated ODDM systems. Similar findings reveal that the PCE-SDF scheme continues to exhibit superior performance compared to the SCE-PDF benchmark scheme during each turbo iteration. Notably, a 1dB gain is observed at a BER of $10^{-4}$ during the first iteration. In addition, for both 4QAM and 16QAM, we can observe that one turbo iteration PCE can outperform three iterations of SCE, which reduces the number of detection iterations required to achieve the same error performance. Therefore, the overall computation complexity for PCE is also reduced compared to SCE.  \par

Fig. \ref{f:fig12} illustrates the BER performance for a rate-$\frac{1}{2}$ LDPC-coded 16QAM modulated ODDM system across different iterations. From the plot, it can be observed that both MMSE-based joint detection and decoding schemes converge in just two iterations, whereas the MRC counterpart requires approximately six iterations to converge. Furthermore, after convergence, the ultimate BER performance aligns with the results achieved by our proposed PCE scheme, demonstrating its effectiveness. \par

\section{Conclusion}\label{sec:con}
In this paper, we proposed a parallel channel encoding (PCE) scheme and derived the soft interference cancellation-based SIC-MMSE detection method for ZP-ODDM modulation. Building on the parallel encoding framework, we also introduced a successive decoding feedback (SDF) iterative detection and decoding algorithm, enabling a closer interaction between the detection and decoding modules. Our analytical and simulation results demonstrate that the proposed PCE-SDF scheme, despite employing multiple shorter codewords, achieves significantly improved error performance compared to the benchmark scheme based on a single long codeword encoding approach.


\bibliographystyle{IEEEtran}
\bibliography{references_form_men,ref}

\vfill

\end{document}